# Comment on "A novel two-dimensional boron–carbon–nitride (BCN) monolayer: A first-principles insight [J. Appl. Phys. 2021, 130, 114301]"

Bohayra Mortazavi[a,*], Fazel Shojaei[b] and Mehdi Neek-Amal[c]

[a]*Department of Mathematics and Physics, Leibniz Universität Hannover, 30167 Hannover, Germany.*
[b]*Faculty of Nano and Bioscience and Technology, Persian Gulf University, Bushehr 75169, Iran.*
[c]*Department of Physics, Shahid Rajaee Teacher Training University, 16875-163 Lavizan, Tehran, Iran*

*Corresponding author: bohayra.mortazavi@gmail.com;

*Abstract*

Recently, Bafekry *et al.* [*J. Appl. Phys. 130*, 114301(**2021**)] reported their density functional theory (DFT) results on the structural, phonon dispersion, elastic constants and electronic properties of a BCN monolayer. The aforementioned theoretical work however includes erroneous results and discussions, as will be explained in the following.

Before discussing the major errors, we first briefly describe the crystal structure of the BCN monolayers considered by Bafekry *et al.*[1]. In their paper, they considered 12 configurations for the BCN monolayers, limiting themselves to hexagonal primitive cells with 2: C, 2: N, and 2: B atoms arranged in graphene-like honeycomb lattices. However, by a careful examination of the structures illustrated in Fig. 1 of their manuscript, it appears that only 5 of these 12 configurations are unique and the rest can be simply generated by rotating those configurations (Configurations (1), (9), (12); configurations (2), (11), (10), configurations (3), (6) and (7), and configurations (4) and (5) are identical). Fig. 1 shows the crystal structure of the energetically most stable lattice, so called configuration (9). This configuration can be thought of as an armchair-shaped [B-N] chains held together by isolated C-C bonds. Nonetheless, they found that the configuration (9) is the most stable structure, whereas it should basically exhibit the same energy as those of (1) and (12) counterparts. The reason that Bafekry *et al.*[1] found different energies for the same systems can be attributed to the fact that their employed criteria for the geometry optimization is rather coarse (0.05 eV/Å). In addition, maybe the applied vacuum length along the sheets' normal direction was rather short (which is not mentioned in their work), resulting in artificial interactions between the layers and change of the calculated energies.

Our DFT calculations are performed by employing the *Vienna Ab-initio Simulation Package* [2,3]. The generalized gradient approximation (GGA) is adopted with the Perdew-Burke-Ernzerhof (PBE) exchange–correlation functional. The plane wave and self-consistent loop cutoff energies are set to 500 and $10^{-6}$ eV, respectively. These simulation setups and the utilized software are exactly the same as those employed in the work by Bafekry *et al.*[1], however the self-consistent loop cutoff energy in their work was set to $10^{-5}$ eV. To obtain the geometry optimized and stress-free structures, atomic positions and lattice sizes are relaxed using the conjugate gradient algorithm until Hellman-Feynman forces drop below 0.001 eV/Å [4] using a 15×15×1 Monkhorst-Pack [5] K-point grid. The periodic boundary conditions are considered in all directions with a 20 Å vacuum distance along the monolayer's thickness. The hexagonal lattice constant of BCN monolayer is found to be 4.347 Å, in a close agreement with that reported in the work by Bafekry *et al.*[1]. Bafekry *et al.*[1] discussed the nature of chemical bonds in BCN monolayer by computing the electron localization function (ELF) [6]. There are two issues with their ELF plot and the corresponding explanation. First, they did not provide a color scale in Figure 2a. Second, they erroneously wrote that "*ELF takes values between 0 and 0.5, where ELF= 0.5 indicates perfect electron localization*". By definition ELF is a dimensionless function that takes a value between 0 to 1. We also calculated the ELF for the BCN monolayer within its primitive cell, as shown in Fig. 1. The large ELF values over 0.8 around the center of bonds indicates the formation of strong covalent interactions throughout the BCN monolayer. Bafekry *et al.*[1] also mentioned that "*electron localization accumulates on N atoms*", which is in a sharp contrast with the ELF result, which shows localizations around the center of bonds and not over individual atoms.

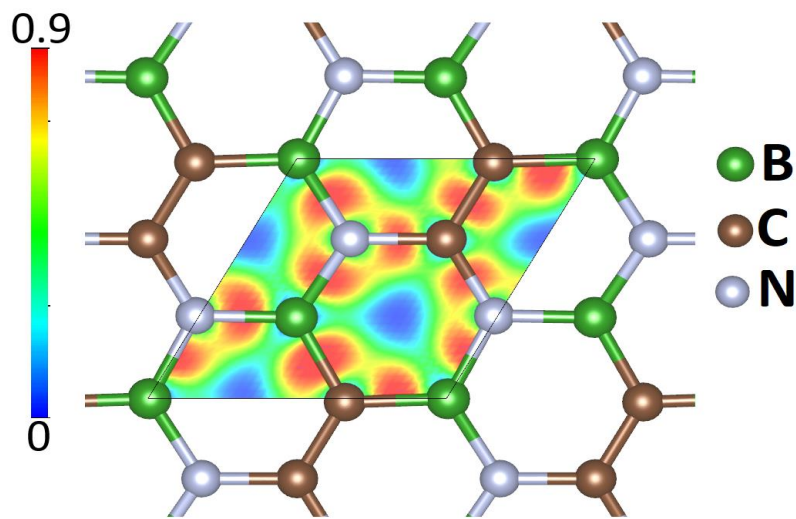

**Fig. 1**, Top view of BCN monolayer along with electron localization function (ELF).

Bafekry *et al.*[1] predicted the Young's modulus of BCN monolayer as 69.3 GPa, which is remarkably low for carbon-based materials, well-known for their superior mechanical properties. Using the DFT method and with assuming a thickness of 3.35 Å for the BCN monolayer as that of the graphene, $C_{11}$, $C_{22}$ and $C_{12}$ of the BCN monolayer are predicted to be 880, 891 and 182 GPa, respectively, equivalent with Young's modulus of 843 and 853 GPa, along the armchair and zigzag directions, respectively. Moreover, Bafekry *et al.*[1] predicted $C_{13}$ and $C_{33}$ of BCN monolayer to be 1.8 and 7.2 GPa, respectively. We note that a monolayer that is free to move toward the out-of-plane direction, cannot exhibit non-zero $C_{13}$ and $C_{33}$ values. This might be once again the outcome of an inaccurate geometry optimization or artificial image-image interactions along the monolayer's normal direction.

Bafekry *et al.*[1] presented their results for the phonon dispersion of the BCN monolayer in Fig. 3a of their manuscript, which also includes two major errors. First, the frequency range extends almost up to 1500 THz, which is clearly wrong. Moreover, symmetrical points of the first Brillouin zone are defined incorrectly, resulting in unsmooth dispersions at K point. In this work we conducted density functional perturbation theory calculations using the VASP package to obtain phonon dispersions over a 4×4×1 supercell utilizing the PHONOPY code [7]. The obtained phonon dispersion of the BCN monolayer is presented in Fig. 2.

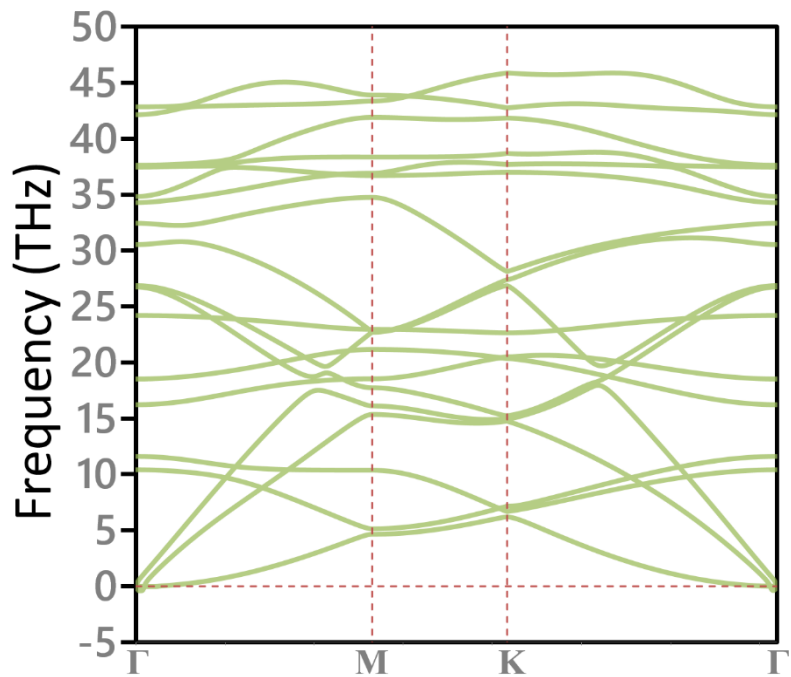

**Fig. 2**, Phonon dispersion relation of the BCN monolayer.

Based on the above discussions, the work by Bafekry et al.[1] is an erroneous study and includes major errors in the structural analysis and results and discussions related to the electron localization function, elastic constants and phonon dispersion relation of the BCN monolayer.

**Acknowledgments**

B.M. appreciates the funding by the Deutsche Forschungsgemeinschaft (DFG, German Research Foundation) under Germany's Excellence Strategy within the Cluster of Excellence PhoenixD (EXC 2122, Project ID 390833453). F.S. thanks the Persian Gulf University Research Council, Iran for support of this study. B. M is greatly thankful to the VEGAS cluster at Bauhaus University of Weimar for providing the computational resources.